\documentclass[prl,twocolumn,english,aps]{revtex4-1}
\usepackage[T1]{fontenc}
\usepackage[latin9]{inputenc}
\usepackage{bm}
\usepackage{amsmath}
\usepackage{graphicx}
\usepackage{amssymb}
\usepackage{esint}

\makeatletter

\providecommand{\tabularnewline}{\\}

\@ifundefined{textcolor}{}
{%
 \definecolor{BLACK}{gray}{0}
 \definecolor{WHITE}{gray}{1}
 \definecolor{RED}{rgb}{1,0,0}
 \definecolor{GREEN}{rgb}{0,1,0}
 \definecolor{BLUE}{rgb}{0,0,1}
 \definecolor{CYAN}{cmyk}{1,0,0,0}
 \definecolor{MAGENTA}{cmyk}{0,1,0,0}
 \definecolor{YELLOW}{cmyk}{0,0,1,0}
 }

\makeatother

\usepackage{babel}

\begin{document}

\title{Sound-Based Analogue of Cavity Quantum Electrodynamics in Silicon}

\author{Ö. O. Soykal, Rusko Ruskov,  and Charles Tahan}


\affiliation{Laboratory for Physical Sciences, 8050 Greenmead Dr., College Park, MD 20740}
\begin{abstract}
A quantum mechanical superposition of a long-lived, localized phonon
and a matter excitation is described.
We identify a realization in strained silicon:
a low-lying donor transition (P or Li)
driven solely by acoustic phonons at wavelengths where high-$Q$    
phonon cavities can be built.
This phonon-matter resonance is shown to enter the strongly-coupled regime where the
``vacuum'' Rabi frequency exceeds the spontaneous phonon
emission into non-cavity modes, phonon leakage from the cavity, and
phonon anharmonicity and scattering.
We introduce a  micropillar distributed Bragg reflector Si/Ge cavity,
where $Q \simeq 10^5-10^6$ and mode volumes   
${\cal V} \lesssim 25\lambda^3$ are reachable.
These results indicate that single or many-body   
devices based on these systems are experimentally realizable.

\end{abstract}

\maketitle

Cavity-quantum electrodynamics (cQED) refers to the
interaction of a single-mode of the electromagnetic field with a
dipole emitter.
cQED has provided new ways of controlling photons and matter (atoms, qubits, etc.) in both atomic and solid-state systems.
The progression from an atom ``dressed'' with a cavity photon (``traditional'' cavity-QED)
\cite{Kimble:1998p1094,raimond:2001}
to a semiconductor microcavity-polariton (exciton plus cavity-photon) \cite{Yamamoto:2010p1032}
to solid-state many-body polaritonic devices \cite{Littlewood:2007p1089,Greentree:2006p745,Kavokin:2003book}
has opened up new avenues for physical investigation
as well as technology
(e.g., single photon sources, novel lasers, long-range entanglement, quantum simulation).
Motivated by this, we seek an analogous progression utilizing quantum
sound instead of light.

Phonons are more suitable for some tasks than photons due to their slower speeds
and smaller wavelengths (e.g., in signal processing, sensing, or nanoscale imaging).
Our work builds off recent experimental results in nano-optomechanical systems \cite{NanoMech-Structures},
where cooling, coherent control, and lasing of mechanical vibrations has been achieved,
as well as previous consideration of phonons
as decoherence pathways \cite{decoher-path},
as tools for coupling quantum systems
\cite{Cirac:1995p1097,Smelyanskiy:2005p860,Rabl:2009p041302},
and even as a means for simulating many-body dynamics \cite{Porras:2004p940}.

In this Letter, we show that a phonon-based analogue of
the cavity-polariton is possible.
Introducing a suitable high-$Q$  phonon cavity, we calculate
the cavity phonon coupling to a two-level system (TLS) in silicon (Fig.~1)
and also
losses due to spontaneous phonon emission from the donor
into non-cavity modes,
phonon leakage from the cavity,   
and phonon anharmonicity and scattering.
Despite the phonon's dependent nature on its host material and the different (non-dipole)
donor/phonon interaction,
a strong coupling regime can be established, similar to cQED,
where the phonon-TLS states are hybridized.
The result of this mixing of cavity-phonon and matter
excitation we term the cavity-\emph{phoniton} \cite{polaron}.

{\it Implementation.}
Silicon is a promising candidate for constructing
a cavity-phoniton system.
The physics of shallow donors in Si have been understood since 1950s
and experimentally verified, while  transitions between
low energy donor states are known to be driven by acoustic phonons
\cite{Wilson:1961p780}.   
The six-fold degeneracy due to Si's  multi-valley conduction band is lifted
both by applied strain (e.g., due to the lattice mismatch with a substrate)
and the sharp donor potential.
Crucially, in $[001]$ compressively strained-Si (Fig.~1),
the first excited state at zero magnetic field of a phosphorous donor
approaches $\Delta_{v}^{P}\simeq 3.02\,\mbox{meV}$ ($0.73\,\mbox{THz}$):
a so-called excited ``valley'' state. 
(The excited valley state has an $s$-like envelope function like the ground state but opposite parity; 
because of this, valley state relaxation times can be much longer than for charge states).
The
energy splitting implies longitudinal (transverse)   
wavelengths of $\lambda_{l}\approx 12.3\,\mbox{nm}$ ($\lambda_{t}\approx 7.4\,\mbox{nm}$).
For comparison, the energy
splitting to the upper $2p$-like state is $ > 30\,\mbox{meV}$
($\lambda_{2p}\approx 1.2\,\mbox{nm}$), unlikely to be amenable to phonon cavities.
Since the P:Si Bohr radius is  $a_{B}^{*}\lesssim 2.5\, \mbox{nm}$ in the bulk,
$\lambda > a_{B}^{*}$
allows for easier donor placement, avoidance of interface physics,
and bulk-like wave functions.

\begin{figure}[t]  
\begin{centering}
\includegraphics[scale=0.25]{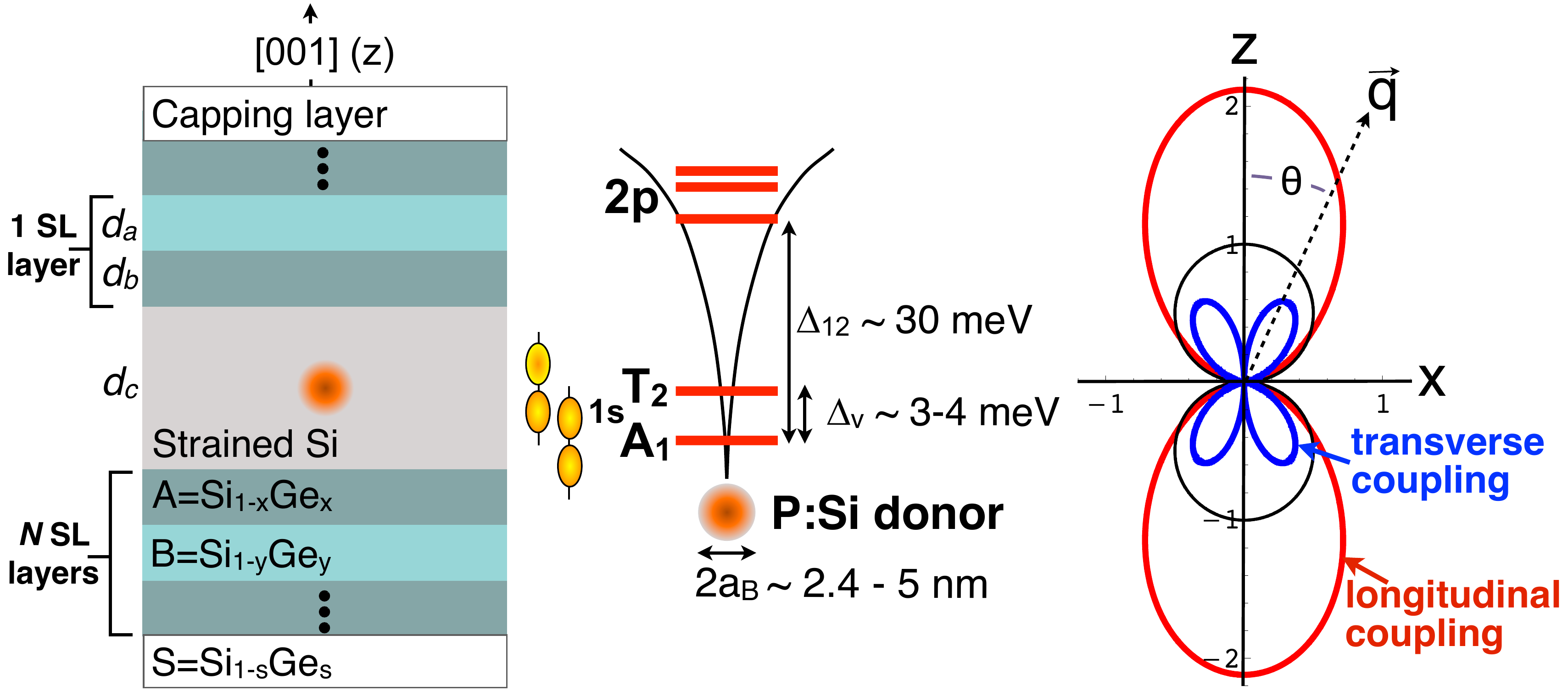}

\par\end{centering}
\caption{
(a) A cavity-phoniton
can be constructed in a Si/Ge heterostructure  cavity
as a hybridized state of a trapped single phonon  mode and a  
donor TLS placed at a maxima of the  phonon field.
(b) The P:Si donor lowest $1s$ valley states, $A_{1}$, $T_{2}$, and upper levels
($1s$/ $2p$); their energy splittings can be controlled by
the applied strain in the Si cavity.
(c) Angular dependence of the
coupling $g_{\bm{q}}(\theta)$, Eq.(\ref{matrix-elements-coupling}),
for the deformation potentials
of Ref.~\cite{YuCardonaBook} vs. dipole $\sim\cos{\theta}$-dependence (thin circles).%
\label{fig:device}
}
\end{figure}

A prototype implementation of a cavity-phoniton system in silicon
is sketched in Fig.~1a.
A strained-Si phonon cavity grown in the $[001]$ direction, of length $d_{c} \sim \lambda$ and lateral size $D$,
is enclosed by acoustic   
DBRs
formed as layered, epitaxially-grown, and strain-relaxed
SiGe super lattice (SL) heterostructures.
The cavity length is chosen to be less than the critical thickness due to strain
(see, e.g., Ref. \cite{Brunner2002}).
The DBR SL unit periods
consist of subsequent layers
$(A,B)=\!\mbox{Si}_{1-x_{A,B}}\mbox{Ge}_{x_{A,B}}$
of thickness ($d_{A},d_{B}$),
strain matched  to a      
$\mbox{Si}_{1-s}\mbox{Ge}_{s}$ substrate
($x_{A(B)}=0.55(0.05),\,s=0.26$, maximizing confinement)
and an appropriate capping
layer, depending on the actually confined phonon mode.
Note that  1D DBR
SL phonon cavities ($D \gg d_{c}$)
are well understood and have been demonstrated
in THz phonon cavities in III-Vs
\cite{NanoMech-Structures,Trigo:2002p922};
coherent phonons in SiGe superlattices  were studied as well \cite{Ezzahri:2007p1057}.
In the case of micropillar DBR (mpDBR) structures
(designed   
to increase the phonon-donor coupling),
the DBR lateral dimension may become comparable to the
phonon wavelength ($D \gtrsim d_{c} \sim \lambda$) and the
confined mode is a mixed longitudinal/transverse one.
%
The trapped mode with wavelength $\lambda_{q}$ and phase velocity $v_{q}$
is designed to be resonant in energy  with the first excited state of the donor.  
Similar to the 1D DBR \cite{Rytov}, the thickness of the SL unit cell
is set to match the Bragg
condition $d^{(q)}_{A,B}=v^{(q)}_{A,B}\lambda_{q}/4v_{q}$,
where $v^{(q)}_{A,B}$ are the phase velocities   
(using isotropic approximation,  
see, e.g., Ref. \cite{Komirenko:2000p1205}).
The donor is placed at the center of  a  
$\lambda$-cavity ($d_{c}=\lambda_q$),
where the  
displacement  ${\bm u}(\bm{r})$ is maximal \cite{zero-strain}.

{\it Hamiltonian and coupling.}
In the semi-classical picture an acoustic phonon creates a time-dependent
strain,
$\varepsilon_{\alpha\beta}(\bm{r})=
\frac{1}{2}\left(\frac{\partial u_{\alpha}}{\partial r_{\beta}}+\frac{\partial u_{\beta}}{\partial r_{\alpha}}\right)$,
which modulates the energy bands
and can drive transitions in
a localized state, e.g., a donor.
For Si, from the multivalley electron-phonon interaction \cite{BirPikusBook,YuCardonaBook}
one can derive the matrix element between valley states, $|s,j\rangle$:
\begin{eqnarray}
&& V_{ij}^{s's} \equiv \hbar g_{\bm{q}} =
i \langle s',i|\, \Xi_d\, \mbox{Tr}(\varepsilon_{\alpha\beta})
\nonumber \\
&&
\qquad\qquad\quad {} + \frac{1}{2} \Xi_u\, \left\{ \bm{\hat{k}}^{\alpha}_i \bm{\hat{k}}^{\beta}_i +
\bm{\hat{k}}^{\alpha}_j \bm{\hat{k}}^{\beta}_j \right\} \varepsilon_{\alpha\beta}\, |s,j\rangle
\label{general-ME} ,
\end{eqnarray}
where $\bm{\hat{k}}_{i,j}$ are the directions toward the valleys,
$s,s'$ label the orbital (envelope) function(s),
and $\Xi_{u}(\mbox{Si})\simeq 8.77\,\mbox{eV}$,
$\Xi_{d}(\mbox{Si})\approx 5\,\mbox{eV}$
\cite{YuCardonaBook}
are deformation potential constants.
For the donor-phonon Hamiltonian
we obtain the interaction
(of Jaynes-Cummings type):
$H_{\rm g}\approx \hbar g_{\bm{q}}\left(\sigma_{ge}^{+}b_{\bm{q},\sigma}+\sigma_{ge}^{-}b_{\bm{q},\sigma}^{\dagger}\right)$
where
only the resonant cavity phonon with
quantum numbers ${\bm{q},\sigma}$ and energy $\hbar\omega_{\bm{q},\sigma}$
is retained, $b_{\bm{q},\sigma}^{\dagger}$ is the phonon creation operator,
and $\sigma_{ge}^{+}\equiv|e\rangle\langle g|$
refers to the donor transition between ground and excited states.
In the loss part:
$H_{\rm loss} = H_{\kappa} + H'_{\rm anh} + H_{\Gamma}$,
$H_{\kappa}$ couples  the cavity mode to
external continuum of other modes giving a cavity  decay rate
 $\kappa=\omega_{\bm{q},\sigma}/Q$
(expressed through the  $Q$-factor);
$H'_{\rm anh}$ includes phonon decay
due to phonon self-interaction and also phonon scattering off
impurities (mainly mass fluctuations in natural Si).
The coupling of the donor to modes other than the cavity mode, $H_{\Gamma}$,
leads to its spontaneous decay.   

The valley states: $1s(A_{1})$, $1s(T_{2})$, that make up the TLS
are the symmetric and anti-symmetric combinations of the conduction
band valley minima   
along the $\hat{z}$-direction (see Fig. 1b).
Due to opposite parity of the states the intravalley contributions cancel.
The intervalley transitions are preferentially
driven by Umklapp phonons \cite{Castner:1963p857}
with a wave vector $\bm{q}$
at $q_{u}\simeq 0.3\,\frac{2\pi}{a_{0}}$, where $\bm{q}_{u}\equiv\bm{G}_{+1}-2\bm{k}_{\hat{z}}$
is the wave vector  ``deficiency'' of the intervalley $\bm{k}_{\hat{z}}\to-\bm{k}_{\hat{z}}$
transition, 
and $\bm{G}_{+1}=\frac{4\pi}{a_{0}}(0,0,1)$ is the reciprocal vector
along $\hat{z}$.
Since typical values give $q_{u}r \approx q_{u}a_{B}^{*}\simeq 9.4 > 1$
and $qr\sim1$
for $3\,\mbox{meV}$, the
coupling is calculated exactly
 (not using the dipole approximation)
for longitudinal and transverse polarizations
\begin{eqnarray}
 &  & g_{\mathbf{q}}^{(\sigma)} = \left(\frac{a_{G}^2\, q^2}{2\rho\hbar{\cal V}\omega_{\bm{q},\lambda}}\right)^{\! 1/2}
 \!\!\! I^{ge}(\theta)
 \left\{
 \begin{array}{c}
\!\Xi_{d}+\Xi_{u}\cos^{2}{\!\theta} \,\,\,\, [\rm l]\\
\!\Xi_{u}\sin{\theta}\cos{\theta} \,\,\,\, [\rm t]
\end{array}\right.
\label{matrix-elements-coupling},
\end{eqnarray}
where $a_{G}\approx 0.3$,
 $I^{ge}(\theta) = \int d\bm{r}[\Phi_{1s}^{\hat{z}}(\bm{r})]^{2}e^{-i\bm{q}\bm{r}}\sin{(\bm{q}_{u}\bm{r})}
 = \frac{2\beta_{q}\cos{\theta}\,(1-\gamma_{q}\cos^{2}{\!\theta})}
 {\alpha_{q}^{2}\,\left[(1-\gamma_{q}\cos^{2}{\!\theta})^{2}-\beta_{q}^{2}\cos^{2}{\!\theta}\right]^{2}}$
is the intervalley overlapping factor,
 $\alpha_{q}=1+\frac{1}{4}(q^{2}a^{2}+q_{u}^{2}b^{2})$, $\beta_{q}=\frac{1}{2\alpha_{q}}b^{2}qq_{u}$,
$\gamma_{q}=\frac{1}{4\alpha_{q}}(a^{2}-b^{2})q^{2}$, and $a/b$
are the radii   
of the Kohn-Luttinger envelope function $\Phi_{1s}^{\hat{z}}(\bm{r})$
(see, e.g. Ref.\cite{YuCardonaBook}).
%
The calculated coupling is to plane wave modes, related to a rectangular
cavity with periodic boundary conditions.
Fig.~1c shows the directionality of the coupling for longitudinal and transverse phonons.
The angular dependence in Eq.~(\ref{matrix-elements-coupling}) for
longitudinal phonons is similar to dipole emission ($I_{\rm dip}^{ge}\sim \cos{\theta}$),
but  
enhanced in a cone around $\hat{z}$-direction  
due to non-dipole contributions.
Uncertainties in  $\Xi_{d}$ calculations (see Ref.~\cite{Leu:2008p1153}) 
result in an overall factor of 2 difference in the maximal coupling.

{\it Coupling to cavity mode.}
The matrix element $\hbar g_{\mathbf{q}}^{(\sigma)}$ is just the
interaction energy of the donor with the
``phonon vacuum field''
and expresses a generic dependence $\propto1/\sqrt{{\cal V}}$ on
the normalization volume;
it goes to zero for large volumes.
In a cavity,  ${\cal V}$ is the physical volume of the mode \cite{Berman1994}.
By virtue of Eq.~(\ref{matrix-elements-coupling}),
we first consider a DBR cavity
with a length $d_{c}=\lambda_{\rm l}\simeq 12.3\,\mbox{nm}$
designed for longitudinal resonant phonon along the $z$-direction
(we use isotropic velocities  for Si:
$v_{l}=8.99\,10^{3}\,\mbox{m s}^{-1}$, $v_{t}=5.4\,10^{3}\,\mbox{m s}^{-1}$, see Ref.\cite{Komirenko:2000p1205}).
Taking the minimal lateral size $D_{\rm min}\approx\lambda_{l}$
to ensure $D > 2 a_{B}^{*}\approx 5\,\mbox{nm}$, one gets a mode volume
of ${\cal V}_{\rm min} \approx \lambda^{3}$ for P:Si.
Thus, we estimate the maximal
phonon-donor coupling as $g_{1\lambda}=3.7\,\,10^{9}\,\mbox{s}^{-1}$.
For $D=5\lambda$ the coupling is still appreciable: $g_{5\lambda}=7.4\,10^{8}\,\mbox{s}^{-1}$.
Surface undulation typical of step-graded SiGe quantum wells (see Ref.~\cite{Brunner2002})
gives $D\approx 200\,\mbox{nm}\lesssim 15 \lambda$, though this interface imperfection is avoidable with heterostructures grown on defect-free nanomembrane substrates \cite{Paskiewicz}.

For a  realistic cylindrical mpDBR cavity,
the modes can be constructed as standing waves with energy $\hbar \omega$, and
wave number $q$ along the pillar $z$-direction,
with
stress-free boundary conditions on the cylindrical surface.
The displacements for compressional modes
(see e.g. Ref.\cite{Clelandbook2003}) are:
$u_r(r,z) = [A_r J_1(\eta_l r) + B_r J_1(\eta_t r)] \sin{qz}$,
$u_z(r,z) = [A_z J_0(\eta_l r) + B_z J_0(\eta_t r)] \cos{qz}$,
where
$J_{0,1}(r)$ are Bessels of 1st kind,
$\eta_{l,t}\equiv \sqrt{ \omega^2 /v_{l,t}^2 - q^2}$,
and $A_i$, $B_i$ are constants.
These modes are lower in energy and couple strongly to the donor;
the related strain has a node at the Si-cavity $z$-boundaries
(for $\lambda$-cavity) \cite{zero-strain}.
For a fixed resonant frequency $\omega$ and lateral size $D$, the dispersion relation $q = \omega/v_q(D)$
has multiple solutions $q_i, i=0,1,2,\ldots$, where $q_0$ stands for the fundamental mode,
$q_1$ for the 1st excited mode, etc.
Each mode propagates with its own phase velocity $v_{q_i}(D)\neq v_{l},v_{t}$.
For a $\lambda$-cavity and $D=\lambda_l$
we calculate via Eq.(\ref{general-ME})  maximal coupling to the fundamental mode
with $\lambda_{q_0}=6.9\, \mbox{nm}$
 to be $g_{1\lambda}^{(0)} = g_{\rm max}=6.5\,\,10^{9}\,\mbox{s}^{-1}$ (see Table I),
comparable to the above estimation.
For larger $D$, however, the coupling to the fundamental mode rapidly decreases
(e.g, $g_{5\lambda}^{(0)}=5.8\,\,10^{5}\,\mbox{s}^{-1}$) since the mode  transforms
to a surface-like  Rayleigh wave  ($v_{\rm Rayleigh} < v_t$).
Coupling to higher mode branches is appreciable and decreases roughly as $1/\sqrt{{\cal V}}$,
e.g., for the 1st mode branch, $g_{1\lambda}^{(1)}=2.4\,\,10^{9}\,\mbox{s}^{-1}$
and $g_{5\lambda}^{(1)}=3\,\,10^{8}\,\mbox{s}^{-1}$.
Among various mode choices we note that for any diameter $D$ there is a higher excited
mode with resonant wavelength close to that for longitudinal phonons.
E.g., for $D=3\lambda_{l}$ we found the wavelength
of the resonant $4^{\rm th}$ excited mode as  $\lambda_{q_4}\simeq 12.4\, \mbox{nm}$
and the coupling  $g_{3\lambda}^{(4)}=3\,\,10^{9}\,\mbox{s}^{-1}$
(i.e., values  similar to the rectangular DBR estimate).

{\it Loss.}
Losses in this system are dominated
by donor relaxation and leakage of the confined phonon mode. Similar
to cQED (see, e.g., Ref. \cite{Berman1994}),
one can argue that the donor relaxation, $\Gamma_{\rm relax}$, to modes
different than the cavity mode (and generally not trapped into the
cavity) is bounded by the donor spontaneous emission rate in the bulk:
$\Gamma_{\rm relax}\lesssim \Gamma$. We calculated the bulk donor
relaxation to longitudinal and transverse phonons to be $\Gamma_{ge}^{(l)}=3\,\, 10^{7}\ s^{-1}$
and $\Gamma_{ge}^{(t)}=9.2\,\, 10^{7}\ s^{-1}$, respectively, for the
$3\,\mbox{meV}$ transition ($\Gamma = \Gamma_{ge}^{(l)}+\Gamma_{ge}^{(t)}=1.2\,\,10^{8}\ s^{-1}$).
The relaxation to photons is electric dipole forbidden
and suppressed \cite{YuCardonaBook} by
 $(\lambda_{photon}/a_{B}^{*})^{2}\sim10^{10}$.

The cavity mode loss rate is calculated as mainly due to leakage through
the DBR mirrors, similar to optical DBR cavites \cite{Pelton:2002p1190}
(except that for phonons there is no leakage through the sides).
Generally, for  the cylindrical micropillar DBRs, the cavity mode
involves coupled propagation along the micropillar of two displacement components, $u_{z}(r,z)$,
$u_{r}(r,z)$, and two stress fields, $T_{zz}$, $T_{zr}$.
For small diameters,  $D \ll\lambda_l$, the fundamental mode becomes
mainly longitudinal ($\sim u_z$),
propagating with the Young velocity, $v_0=\sqrt{E/\rho}$.
Using $4\times 4$ transfer matrices we calculated
$Q_{\rm mpDBR}^{(0)}\simeq \,10^{6}$ for $N=33$ layers for the
confined, mixed fundamental mode at $D=\lambda_{l}$, that is close to
the limiting $Q$-factor related to pure longitudinal propagation\cite{1D-DBR}.
(This is also similar to the 1D DBR value
relevant for $D\gg\lambda_l$).
For our design we
obtain a cavity loss rate $\kappa=\Delta_{v}/\hbar Q\simeq 2.8\,\, 10^{6}\, s^{-1}$.
This can be decreased by adding more layers.

At low temperatures the phonon anharmonicity losses
are negligible (a rate $\Gamma_{\rm anh}\simeq 1.4\,\, 10^{4}\,\mbox{s}^{-1}$
at $3\,\mbox{meV}$), while scattering off impurity mass fluctuations
in natural Si amounts to a rate two orders of magnitude larger: $\Gamma_{\rm imp}\approx 7\,10^{5}\,\mbox{s}^{-1}$.
It is notable that in isotopically purified bulk silicon
(an enrichment of $^{28}\mbox{Si}$ to 99\%) the scattering rate will decrease by
an order of magnitude and the related phonon mean free path will be
of the order of $v_{l}/\Gamma_{\rm imp}\approx 10\,\mbox{cm}$
\cite{Schwab:2000p1289,Clelandbook2003,roughness}.
In this case, the cavity leakage dominates, $\kappa\gg\{\Gamma_{\rm anh},\Gamma_{\rm imp}\}$,
and the number of vacuum Rabi flops can reach as high as
$n_{\rm Rabi} = 2g(D)/(\Gamma + \kappa)\simeq 102$ for
a cavity $Q$-factor, $Q = 10^{6}$, and some $n_{\rm Rabi}\simeq 77$
for $Q=10^{5}$.
For $D=10\lambda_l$ and similar $Q$ one still has:
$n_{\rm Rabi}^{(1)} \simeq 1\, (17)$
for the 1st (2nd) excited mode.
Further, nearby modes can be well separated from the resonant mode, e.g.
for the fundamental mode and $D = \lambda_{l}$ the next mode
(in transverse direction) is $\sim 0.3\, \Delta_v = 0.9\,\mbox{meV}$ off;
the transverse separation for $D = 10 \lambda_{l}$ (for the 1st excited mode) gives
$0.009\, \Delta_v$,
which is more than two orders of magnitude larger than the linewidth
$\Gamma_{0}\simeq(\Gamma+\kappa)/2$ of the two hybridized levels.

\begin{table}
{\footnotesize
\begin{tabular}{|l|l|c|c|c|}
\hline
parameter  & symbol  &  circuit-QED  & {\scriptsize P:Si}/phonon & {\scriptsize Li:Si}/phonon \tabularnewline
\hline
resonance freq.  &\, $\omega_{\rm {r}}/2\pi$  &  $10\,{\rm {GHz}}$  & $730\,{\rm {GHz}}$  & $142\,{\rm {GHz}}$ \tabularnewline
\hline
vac. Rabi freq.  &\, $g/\pi$       &  $100\,{\rm MHz}$  & $2.1\,{\rm GHz}$  & $13.8\,{\rm MHz}$ \tabularnewline
\hline
cavity lifetime  &\, $1/\kappa,Q$  &  $160\,{\rm ns}$, $10^{4}$  & $ 0.22\,{\rm \mu s}$, $10^{6}$  & $1.1\,{\rm \mu s}$, $10^{6}$  \tabularnewline
\hline
TLS lifetime  &\, $1/\Gamma$  &  $2\,{\rm \mu s}$  & $8.2\,{\rm {ns}}$  & $22\,{\rm \mu s}$ \tabularnewline
\hline
critical atom {\scriptsize\#}  &\, $2\Gamma\kappa/g^{2}$  &  $\lesssim6\,10^{-5}$  & $\lesssim 3\,10^{-5}$  & $\lesssim 4\,10^{-5}$  \tabularnewline
\hline                                                                                                               
crit.\! phonon {\scriptsize\#}  &\, $\Gamma^{2}/2g^{2}$  &  $\lesssim \,10^{-6}$  & $\lesssim 2\,10^{-4}$  & $\lesssim 6\,10^{-7}$  \tabularnewline
\hline
{\scriptsize\#} Rabi flops  & ${\scriptsize 2g/(\!\kappa\!+\!\Gamma)}$  &  $\sim 100 $  & $\sim 102$  & $\sim 93$  \tabularnewline
\hline
\end{tabular}
}
\caption{
Key rates and 
parameters for
circuit QED \cite{Blais:2004p1107}
(1D cavity) vs. the phoniton system;
we show calculations for maximal coupling
for a $\lambda$-cavity
with lateral diameter of $D=\lambda_l$
(in general, $g \sim 1/\sqrt{\cal V}$),
$Q=10^{6}$, and comparable number of Rabi flops \cite{transit_time}.
\label{tab:Key-rates}
}

\label{table:parameters}
\end{table}

{\it Experiment and Discussion.}
We have shown that the donor:Si  cavity-phoniton
can enter the strong
coupling regime with $2g/(\Gamma+\kappa)\sim 10-100$ (Table \ref{tab:Key-rates}).
A principle experimental confirmation
would be the observation of the
``vacuum'' Rabi splitting:
$\Omega_{0} = \left[g^{2}-(\Gamma-\kappa)^{2}/4\right]^{1/2}$;
two resolved spectral peaks
can be observed if $2\Omega_{0} > \Gamma_{0}$.  
%
The Rabi splitting can be enhanced as $\Omega_{0}\simeq g\sqrt{N}$ by
placing more than one donor ($N > 1$)
in the cavity  \cite{Kimble:1998p1094,Berman1994} (e.g., via a delta-doped layer).
This could
allow for large coupling   
even for large diameter micropillars/
1D DBR structures
(since $\frac{\sqrt{N}}{\sqrt{\cal V}}\propto\frac{\sqrt{D^{2}}}{D} = 1$).
Further, strain  or 
electric field (from a top gate) \cite{Smelyanskiy:2005p860,Friesen:2005p1104,transport}
can be used to tune the valley transition into
resonance.

Experimental techniques are available to probe the Si-phoniton
(low temperature, $T \sim 1\,\mbox{K}$, and low phonon numbers are assumed).
First, free-electron lasers
have been used to probe the $1s-2p$ transitions in P:Si
\cite{Greenland:2010p904}.
Observation of the vacuum Rabi splitting is possible by measurement of the absorption
spectrum of the allowed optically probed transition $1s(T_{2})\rightarrow 2p_{0}$
($\sim 30\,\mbox{meV}$) using weak optical excitation.
Appropriate  phonons can be introduced to the system
by excited valley state emission,
by piezo-actuators, or
by increasing the temperature, as
was done for the first observations of Rabi oscillations \cite{Rempe:1987p1130}
(phonons of $3 \mbox{ meV}\sim 30\,\mbox{K}$).
Second, pump and probe optical techniques have been demonstrated
to observe coherent phonon effects in III-V   
\cite{NanoMech-Structures,Trigo:2002p922}
and SiGe \cite{Ezzahri:2007p1057}   
SL heterostructures.
Observing the reflected phonons from this structure
will show the phonon-Rabi splitting characteristic of cavity-QED
systems \cite{Kimble:1998p1094}.

The cavity-phoniton can be realized    
in other materials and systems.
In particular, our system should be compatible with
recently demonstrated (though in the few GHz range)
high-Q phononic band-gap    
nano-mechanical and opto-mechanical (membrane) cavities in silicon (e.g., in 
\cite{NanoMech-Structures} and \cite{SafaviNaeini:2010p1193}).   
%
%
Quantum dots, spin transitions, color centers in diamond, 
and other donors
(particularly Li:Si \cite{Smelyanskiy:2005p860})
may offer smaller resonance energies
and correspondingly larger cavities (wavelengths) \cite{roughness}.
In the case of $[001]$-strained Si considered in this paper,
the two lowest levels in Li possess essentially the same state structure as P:Si
and approach a splitting of $\Delta_{\rm Li} = 0.586\,\mbox{meV}$ for high strain
(from zero splitting at no strain).
We calculate (see Table \ref{tab:Key-rates})
the corresponding Li:Si donor-phonon coupling 
for the $D=\lambda_{l}$  (now $\lambda_{l}=63.2\,\mbox{nm}$) reference cavity; strong coupling can still be reached.
For the DBR cavities, the Si critical
thickness can be made $80-100\, \mbox{nm}$ by lowering the Ge content
in the substrate.
For 2D-phononic bandgap cavities,
direct numerical calculations for cavity-trapped phonons
($\omega_r/2\pi \approx 10\,\mbox{GHz}$)
in novel Si-nanostructures \cite{SafaviNaeini:2010p1193} show the
potential to reach
$Q_{\rm cav}\gtrsim 10^{7}$ in the ideal case.

The phoniton is a new component for constructing and controlling macroscopic artificial
quantum systems based on sound.
Besides single phonon devices, systems composed of many coupled phonitons could exhibit novel
quantum many-body behavior.
For example, ``solid-sound''
systems in analogy with coupled cavity-QED solid-light systems \cite{Greentree:2006p745}
could demonstrate Mott insulator like states of phonons in coupled
phoniton cavities.
Cavity/qubit
geometries such as these may also be relevant for quantum computing (QC)
applications: to mediate interactions between distant qubits or inhibit
decoherence.
The systems proposed here will
benefit from the drive in silicon QC towards more purified materials, perfect surfaces,
and precise donor placement.

{\it Acknowledgments.}
Special thanks to Chris Richardson for input on realistic SiGe heterostructures
and to Ari Mizel, Robert Joynt, and Mark Friesen for
valuable discussion.
This work was funded in part by DARPA.

\clearpage
\newpage

\section*{SUPPLEMENTAL MATERIAL}

\subsection*{Deriving the phoniton Hamiltonian}

In semiconductors at low energies, the conduction band plays a role
similar to the QED vacuum: the quasi-free electrons propagate according
to a quadratic dispersion, scatter off acoustic phonons that possess
linear dispersion, and can be bound to charged impurity potentials
to establish hydrogen-like atoms (donors).
Thus, the pattern of QED (matter and photon interactions) repeats
itself in the behavior of electronic quasiparticles in the solid (the
{}``matter'') and the phonon vibrations in a crystal (the {}``photons'').
Consider the electron-phonon interaction. The Hamiltonian can be rearranged
into the electron part $H_{\rm e}$, the phonon part $H_{\rm ph}$,
and the electron-phonon interaction $H_{\rm e,ph}$: $H=H_{\rm e}+H_{\rm ph}+H_{\rm e,ph}$,
where $H_{\rm e}=\frac{p^{2}}{2m}+V_{\rm lat}(\bm{r})$ includes
the periodic lattice potential $V_{\rm lat}(\bm{r})$ of the perfect
crystal acting on an electron at point $\bm{r}$. For small atomic
displacements $\bm{u}(\bm{r})$ 
an expansion in normal modes gives
\begin{equation}
\bm{u}(\bm{r})=\sum_{\bm{q},\sigma}\left(\bm{u}_{\bm{q}\sigma}(\bm{r})\, b_{\bm{q}\sigma}+\bm{u}_{\bm{q}\sigma}^{*}(\bm{r})\, b_{\bm{q}\sigma}^{\dagger}\right),
\label{displacement}
\tag{S1}
\end{equation}
 which approximately diagonalizes the phonon part $H_{\rm ph}$:
$H_{\rm ph}=\sum_{\bm{q},\sigma}\hbar\omega_{\bm{q}\sigma}\left(b_{\bm{q}\sigma}^{\dagger}b_{\bm{q}\sigma}+\frac{1}{2}\right)+H_{\rm anh}$.
The small anharmonicity, $H_{\rm anh}=c\, b_{\bm{q}\sigma}^{\dagger}b_{\bm{q'}\sigma}b_{\bm{k}\sigma}+\cdots$,
is related to phonon self-interaction.
The mode normalization in Eq. \ref{displacement} is 
$\int d^{3}\bm{r}\,\bm{u}_{\bm{q}\sigma}^{*}(\bm{r})\bm{u}_{\bm{q}\sigma}(\bm{r})=\frac{\hbar}{2\rho\,\omega_{\bm{q}\sigma}}$,
so that $b_{\bm{q}\sigma}^{\dagger}$ creates a phonon in the mode
$\bm{q},\sigma$ with energy $\hbar\omega_{\bm{q}\sigma}$ in a material
with mass density $\rho$. The vector $\bm{q}$ denotes a collective
index of the discrete phonon mode defined via the phonon cavity boundary
conditions and quantization volume ${\cal V}$. In particular, the
plane wave expansion corresponds to rectangular periodic boundary conditions and
$\bm{u}_{\bm{q}\sigma}(\bm{r})=\left(\frac{\hbar}{2\rho{\cal V}\,\omega_{\bm{q}\sigma}}\right)^{1/2}\!\!\bm{\xi}_{\bm{q},\sigma}\, e^{-i\bm{q}\cdot\bm{r}}$
with wave vector $\bm{q}$, polarization $\bm{\xi}_{\bm{q},\sigma}$,
and phonon branch $\sigma$. 

By considering low-energy acoustic phonons
 for electrons close to the band minimum, $\bm{k}\approx\bm{k}_{0}$,
the electron-phonon interaction can be 
written as
\begin{equation}
H_{\rm e,ph}^{\rm ac}(\bm{r})=\sum_{ij}D_{ij}\,\varepsilon_{ij}(\bm{r}).
\label{acoustic2}
\tag{S2}
\end{equation}
 The operator $D_{ij}=-\hat{p}_{i}\hat{p}_{j}/m+V_{ij}(\bm{r})$
(with  $\hat{\bm{p}}=-(i/\hbar)\nabla$ and a crystal model dependent $V_{ij}(\bm{r})$)
coincides with the constant-strain deformation potential \cite{HerringVogt,BirPikusBook}
and the strain
$\varepsilon_{ij}(\bm{r})=\frac{1}{2}\left(\frac{\partial u_{i}}{\partial r_{j}}+\frac{\partial u_{j}}{\partial r_{i}}\right)$
causes transitions between electronic states.
A donor impurity center makes a perturbation to the perfect crystal,
thus binding the low-energy electrons. One introduces electron field
operators $\Psi(\bm{r})=\sum_{s}c_{s}\psi_{s}(\bm{r})$ related to
the donor bound states $\psi_{s}(\bm{r})$, where $c_{s}$ is the
annihilation operator for that state.
By integrating Eq.~(\ref{acoustic2}) over the charge density operator,
$\Psi^{\dagger}(\bm{r})\Psi(\bm{r})$, the electron-phonon interaction
takes the second-quantized form:
$H_{\rm e,ph} = \sum_{s,s'}\sum_{\bm{q},\sigma}c_{s'}^{\dagger}c_{s}(b_{\bm{q},\sigma}+b_{\bm{q},\sigma}^{\dagger})\, V_{\bm{q},\sigma}^{s's}$
where the matrix element,
$V_{\bm{q},\sigma}^{s's}=\langle\psi_{s'};\{\bm{q},\sigma\}|H_{{\rm e,ph}}^{{\rm ac}}|\psi_{s}\rangle$,
describes phonon transitions between the bound states.

Similar to quantum optics, the electron-phonon interaction $H_{\rm e,ph}$
describes transitions in an $n$-level atom (although the analog of
electric-dipole approximation is not made). The Jaynes-Cummings type Hamiltonian \cite{Berman1994}
results when a {\it single}
phonon mode $\bm{q}$ is in resonance with the transition (i.e., $\omega_{\bm{q},\sigma}\approx\omega_{ss'}$).
This is possible if we have a suitable phononic cavity and the phononic
bandwidth $\Delta\omega_{\bm{q}}$ is much smaller than the transition
frequency $\omega_{ss'}$. Introducing rising (lowering) operators
$\sigma^{+}=c_{s'}^{\dagger}c_{s}$ ($\sigma^{-}=c_{s}^{\dagger}c_{s'}$)
for the specified transition $(s\to s')$ and supposing real matrix
elements such that $V_{\bm{q},\sigma}^{s's}=V_{\bm{q},\sigma}^{ss'}\equiv\hbar g_{\bm{q}}$,
we obtain
 $H_{\rm g}=\hbar g_{\bm{q}}\left(\sigma^{+}+\sigma^{-}\right)\left(b_{\bm{q},\sigma}+b_{\bm{q},\sigma}^{\dagger}\right)$.
In the typical case of $g_{\bm{q}}\ll\omega_{ss'}$
only the {}``energy conserving'' operators $\sigma^{+}b_{\bm{q},\sigma}$,
$\sigma^{-}b_{\bm{q},\sigma}^{\dagger}$ will survive
in the rotating wave approximation 
obtaining
$H_{\rm g}\approx H^{{\rm JC}}=\hbar g_{\bm{q}}\left(\sigma^{+}b_{\bm{q},\sigma}+\sigma^{-}b_{\bm{q},\sigma}^{\dagger}\right)$.
The total Hamiltonian in such a cavity approximation can then be written as
\begin{align}
H &=\hbar\omega_{\bm{q},\sigma}\left(b_{\bm{q},\sigma}^{\dagger}b_{\bm{q},\sigma}+\frac{1}{2}\right)+\frac{\hbar\Delta}{2}\sigma_{z}+H_{{\rm g}}
 \nonumber \\
  & \qquad{}+H_{\kappa}+H'_{{\rm anh}}+H_{\Gamma},
 \label{total-Hep}
\tag{S3}
\end{align}
%
 where we have denoted the atom transition frequency as $\Delta\equiv\omega_{ss'}$.

\subsection*{Phonon-donor coupling}

For shallow donors the wave function is localized in {}``valleys''
in $k$-space, near the local conduction band minima $\bm{k}_{j}$,
and for Si (tetrahedral symmetry) these are displaced at $k_{0}\simeq 0.85\,2\pi/a_{0}$
along the equivalent directions $\hat{x}$, $-\hat{x}$, $\hat{y}$,
$-\hat{y}$, $\hat{z}$, $-\hat{z}$ in the qubic lattice. The valley
state is a modulated Bloch wave: $\psi_{s}^{j}(\bm{r})\approx\psi_{\bm{k}_{j}}(\bm{r})\Phi_{s}^{j}(\bm{r})$
with the well known envelope functions $\Phi_{s}^{j}(\bm{r})$ \cite{Kohn:1955p1123}.
Each donor bound state is superpositions of such valley states,
$\psi_{s}(\bm{r})=\sum_{j=1}^{6}\alpha_{j}^{s}\psi_{s}^{j}(\bm{r})$
with $\alpha_{j}^{s}$ being the {}``valley populations''. The six-fold
degeneracy of conduction electrons in silicon is broken both by strain
(e.g., due to the lattice mismatch with a Si$_{1-x}$Ge$_{x}$ substrate)
and the sharp potential of the phosphorous (or other) donor. Thus
the two lowest states in our device approach symmetric and antisymmetric
combinations of the $\pm z$ valley states (see Fig. 1, main text) \cite{Wilson:1961p780}
and asymptote to $E_{s's}=E_{e}-E_{g}\simeq3.02\,\mbox{meV}$ splitting
for large ($x\gtrsim 0.1$) mismatch.
Both states are ``s-like'', having the same $s$-like envelope function
but opposite parity due to different content of the highly oscillating Bloch components.
The next highest state is ``$p$-like''
and lies some $30$ meV above the ground state.

One can calculate the partial matrix element for inter-valley transitions
in the long wave-length (acoustic) limit \cite{Kohn:1955p1123,Castner:1963p857}:
\begin{align}
&V_{ij}^{s's}(\bm{q},\sigma) = \left(\frac{\hbar}{2\rho{\cal V}\omega_{\bm{q},\sigma}}\right)^{1/2}
\nonumber \\
&\!\!\! \times\left(\Xi_{d}(\bm{q}\bm{\xi}_{\bm{q},\sigma}) +
\frac{1}{2}\Xi_{u}\left\{ (\bm{q}\hat{\bm{k}}_{i})(\bm{\xi}_{\bm{q},\sigma}\hat{\bm{k}}_{i})+(\bm{q}\hat{\bm{k}}_{j})(\bm{\xi}_{\bm{q},\sigma}\hat{\bm{k}}_{j})\right\} \right)
\nonumber \\
&\!\!\!\times\sum_{\bm{G}}a_{\bm{G}}^{ij}\int d\bm{r}\left[\Phi_{s'}^{i}(\bm{r})\right]^{*}\Phi_{s}^{j}(\bm{r})\ e^{-i(\bm{k}_{i}+\bm{q}+\bm{G}-\bm{k}_{j})\bm{r}}.
\label{Vij-acoustic}
\tag{S4}
\end{align}
 The total matrix element is given by a sum over all valley contributions
weighted by the valley populations: $V_{\bm{q},\sigma}^{s's}=\sum_{i,j}\alpha_{i}^{s'}\alpha_{j}^{s}\, V_{ij}^{s's}(\bm{q},\sigma)$.
Here, $\Xi_{d}$ and $\Xi_{u}$ are the low-$\bm{q}$ deformation
potentials (that compose $D_{ij}$ in Eq. \ref{acoustic2}), $\hat{\bm{k}}_{i}$
is a unit vector towards the $i$th 
valley (e.g. $\hat{\bm{k}}_{1} = \hat{x},\hat{\bm{k}}_{2} = -\hat{x}$, etc.),
and $a_{\bm{G}}^{ij}$ are the Fourier series expansion coefficients
of the function $u_{\bm{k}_{i}}^{*}(\bm{r})u_{\bm{k}_{j}}(\bm{r})$
($u_{\bm{k}}$ are the periodic parts of the Bloch waves), i.e.
$u_{\bm{k}_{i}}^{*}(\bm{r})\, u_{\bm{k}_{j}}(\bm{r})=\sum_{\bm{G}}\, a_{\bm{G}}^{ij}\, e^{-i\bm{G}\bm{r}}$.
In the sum over $\bm{G}$ the main contribution comes from the minimal
reciprocal vectors. Higher terms are suppressed due to spatial confinement
of the orbital part and/or rapid decreasing of the Fourier coefficients
$a_{\bm{G}}^{\hat{z}}$ for higher $\bm{G}$ \cite{Koiller:2004p872}.
Due to different parity of the valley populations under inversion
$\bm{k}_{\hat{z}}\to\bm{k}_{-\hat{z}}$ the intravalley contributions
($i=j$) will cancel. Leading contribution to $V_{\bm{q},\sigma}^{s's}$
comes from intervalley (Umklapp) transitions with $\bm{q}\approx \bm{q}_{u}\equiv \bm{G}_{+1}-2\bm{k}_{\hat{z}}$,
where $\bm{G}_{\pm1}=\frac{4\pi}{a_{0}}(0,0,\pm1)$ is the reciprocal
vector along the $\hat{z}$-direction \cite{Castner:1963p857}.

\subsection*{Donor relaxation}

The total emission rate is calculated from Eq.~(\ref{Vij-acoustic})
via the Golden Rule:
$\Gamma_{ge}^{(\sigma)}=\sum_{\bm{q}}\frac{2\pi}{\hbar^{2}}|V_{\bm{q},\sigma}^{ge}|^{2}\delta(\omega_{\bm{q},\sigma}-\omega_{ge})$.
Using the notations $x\equiv\cos{\theta}$, $J(x;\beta_{q},\gamma_{q})\equiv\frac{\alpha_{q}^{2}}{2\beta_{q}}I^{{\rm ge}}(x)$
(cf. Eq. 2 of the main text),
and
$\Xi^{(\rm l)}(x) \equiv \Xi_{d}+\Xi_{u} x^{2}$, $\Xi^{(\rm t)}(x) \equiv \Xi_{u} x \sqrt{1-x^2}$
we obtain the emission rates for acoustic modes: 
\begin{align}
\Gamma_{ge}^{(\sigma)} &=\frac{(bq_u)^2\omega_{ge}^{5}b^2}{4\pi\rho\hbar\, v_{\sigma}^7}\frac{|a_{G}|^2}{\left[1+\frac{1}{4}\left(q^2 a^2+q_u^2 b^2\right)\right]^6}
\nonumber \\
& {}\times\int_{-1}^{1}dx\,\left.[\Xi^{(\sigma)}(x)\, J(x;\beta_{q},\gamma_{q})]^2\,\right|_{q=q_{ge}^{(\sigma)}},
\label{emission-rate2}
\tag{S5}
\end{align}
 where $v_{\sigma}$ is the speed and $q_{ge}^{(\sigma)}=\omega_{ge}/v_{\sigma}$
is the wave number of the emitted phonon of polarization $\sigma$.
We mention that $\Gamma_{ge}^{(\sigma)}$ as a function of energy
of the transition must experience a maximum when $q_{ge}^{(\sigma)}$
is close to the Umklapp value $q_{u}$ or $\frac{E_{ge}}{\hbar v_{\sigma}}\approx 0.3\, \frac{2\pi}{a_{0}}$.
This corresponds to transition energies $19.2\,\mbox{meV}$ and $13.3\,\mbox{meV}$
for longitudinal and transverse phonons respectively,
which is in accordance with Eq. (\ref{emission-rate2}).
Using the mass density $\rho=2330\,\mbox{kg m}^{-3}$ for Si and the deformational
potential constants: $\Xi_{u}(\mbox{Si})\simeq 8.77\,\mbox{eV}$, $\Xi_{d}(\mbox{Si})\simeq 5.0\,\mbox{eV}$,
and $a_{G}\approx 0.3$ \cite{Koiller:2004p872} one obtains the rates at $3\, \mbox{meV}$: 
\begin{equation}
\Gamma_{ge}^{(l)}=3\,\, 10^{7}\ s^{-1},\ \ \Gamma_{ge}^{(t)}=9.2\,\, 10^{7}\ s^{-1}
\label{rates-3meV}  .
\tag{S6}
\end{equation}
For reference, the values calculated in the dipole approximation are
about $\approx 35-50$\% larger than the above.

\subsection*{Cavity design and cavity quality factors}

\subsubsection*{1D DBR cavity}

\begin{figure*}
\begin{centering}
\includegraphics[scale=0.45]{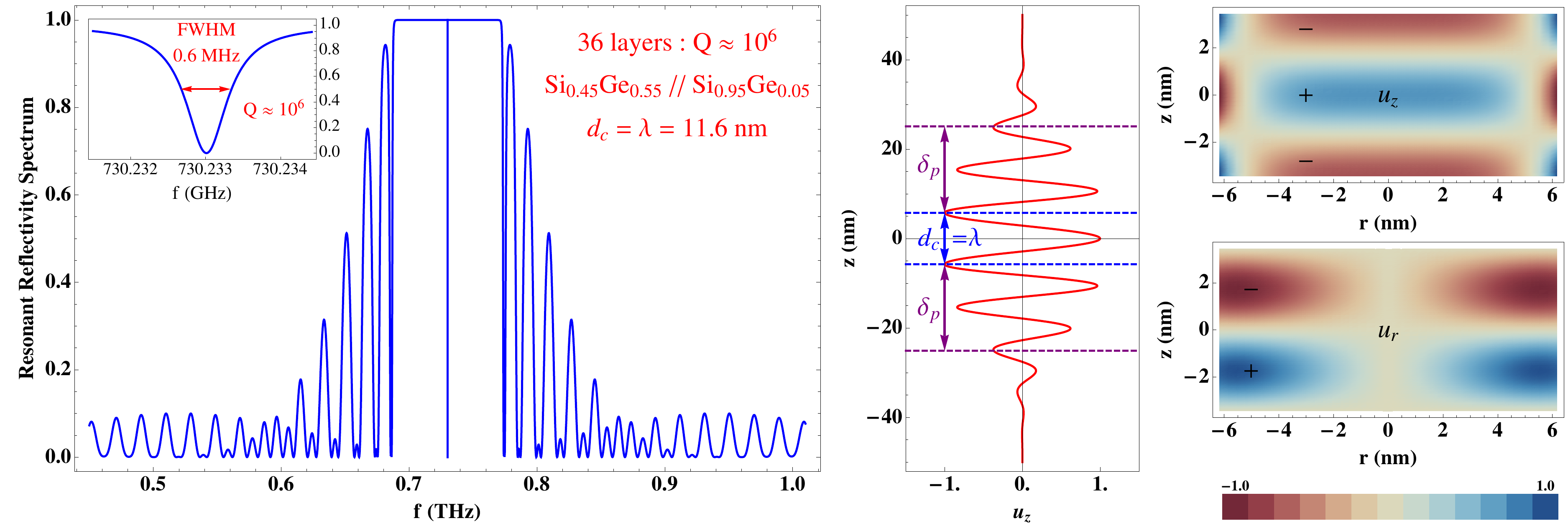}
\par\end{centering}
\caption{
Reflectivity and field distribution of acoustic distributed Bragg reflector (DBR) phonon cavity.
(left) Reflectivity spectrum of the phonon cavity enclosed by DBRs consisting
of $N=36$ subsequent layers of thicknesses $d_A \simeq 2.1\, \mbox{nm}$/$d_B \simeq 2.8\, \mbox{nm}$
($A = \mbox{Si}_{0.45}\mbox{Ge}_{0.55}$/$B = \mbox{Si}_{0.95}\mbox{Ge}_{0.05}$).
Slight thickness changes due to finite interlayer spacing approximation
do not cause significant change in $Q$.
For the 1D calculations we use the [001] phonon velocity
of $v_l = 8.43\, 10^3\, \mbox{\rm m/s}$ giving $\lambda_l \simeq 11.6\, \mbox{nm}$.
The sharp dip in the reflectivity at $f_{0}=0.73\, \mbox{THz}$ corresponds to the donor
transition energy $\sim 3\, \mbox{meV}$.
The left inset shows a zoom in of the reflectivity dip and
corresponds to the confined
cavity mode with FWHM of $0.6\, \mbox{MHz}$ and $Q\approx 10^{6}$.  
(middle)
Amplitude of the localized phonon displacement in the 1D DBR cavity is shown with respect
to the position along $[001]$ in Si,
with the Si cavity of thickness $d_{c}$ and penetration depth $\delta_{p}$ into the DBR, respectively.
(right) 
The mixed phonon mode in a 3D micro-pillar DBR structure.
Normalized amplitudes of the           
phonon displacements $u_z(r,z)$ and $u_r(r,z)$ are shown
for the fundamental mode inside
the Si cavity of thickness $6.9\, \mbox{nm}$  and diameter $D=12.3\, \mbox{nm}$.
The DBR mirrors enclosing the Si cavity are
constructed from alternating layers of Si$_{0.95}$Ge$_{0.05}$/Si$_{0.45}$Ge$_{0.55}$ with thicknesses of $1.7$/$1.2$ nm, respectively.
3D cavity reflectivity spectrums are qualitatively similar to the 1D case.
\label{fig:1Dcavity}
}
\end{figure*}

Due to the analogy between
light and sound in solids, the theory of optical DBRs can be applied to the design
of acoustic DBRs.
For acoustic waves propagating perpendicularly to the 1D DBR SL mirror,
Bragg type interference leads to forming of frequency
stop-bands \cite{Rytov} in the reflectivity spectrum,
positioned around a central frequency (for the $m^{\rm th}$-order band)
$\omega_{m}=m\,\pi \left(d_{A}/v_{A}+d_{B}/v_{B}\right)^{-1}$.
The presence of a resonant Si-cavity (a ``defect'') in the SL leads to a dip in
the reflectivity spectrum
(cf. Fig.~\ref{fig:1Dcavity}),
whose width would represent the 1D DBR $Q$-factor, providing the cavity losses
are mainly through the DBR mirrors.
Similar to optics, the transfer matrix method (see, e.g., Ref.~\cite{Kavokin:2003book})
allows us to evaluate
$Q_{DBR} \simeq 2\pi(d_{c} + L_{DBR})\sqrt{R_c}/\lambda\ln{R_c}$,
where
$L_{DBR} = Z_A Z_B \lambda_q/2 Z_c |Z_B-Z_A|$ is the acoustic effective DBR mirror length
and
$R_{c}=\left[(Z_s/Z_c - Z_{r}^{2N})/(Z_s/Z_c + Z_{r}^{2N})\right]^{2}$
is the DBR peak power reflectivity calculated for $N$ layers with
layer impedance ratio
$Z_{r} = \rho_A v_A/\rho_B v_B$
($Z_s/Z_c$ are the impedances of the substrate/cavity respectively).

The cavity quality factor
extracted numerically from the linewidth of the reflectivity dip (see Fig.~\ref{fig:1Dcavity})
agrees well with the simple analytics.
For our design, we choose the layer materials
$A=\mbox{Si}_{0.45}\mbox{Ge}_{0.55}$  and $B=\mbox{Si}_{0.95}\mbox{Ge}_{0.05}$
with thicknesses of $d_A=2.1\, \mbox{nm}$ and $d_B=2.8\, \mbox{nm}$
(strain matched \cite{Brunner2002} to a substrate of $\mbox{Si}_{0.74}\mbox{Ge}_{0.26}$),
to confine a longitudinal phonon ($d_c=\lambda_l$);
we obtain a $Q$-factor of $\sim 1.2\,\, 10^{5}$ for $N=29$ SL layers;
for $N=36$ $Q_{\rm DBR}$ reaches $10^{6}$.

\subsubsection*{Micro-pillar DBR cavity}

While longitudinal and transverse phonons formally decouple
under isotropic approximation in the bulk, they combine
in  mpDBR cavity/waveguide modes due to the strong effect
of the side boundaries, and propagate with a differing phase velocity $v_q \neq v_l, v_t$.
For the cylindrical mpDBR cavity we focus on compressional modes only,
since flexural modes are energetically higher, while torsional
(transverse) modes do not couple to the donor.
The boundary condition on the mpDBR cavity $z$-interfaces is linked to the  Bragg
resonance condition at the first zone center stop band (for $\lambda_q$-cavity), leading to
$q = 2\pi/d_{c} = \Delta_v/\hbar v_q$,
where $q$ is the cavity phonon wave vector along the pillar $z$-axis.
Additionally, the free-standing (stress-free)
boundary condition along the cylindrical surface implies for the stresses:
$T_{rr}=T_{rz}=0$ at $r=R$,
and leads to the well-known Pochhammer-Chree dispersion relation of
compressional waves (see, e.g., Ref. \cite{Clelandbook2003}),
$2\eta_{l}(\eta_{t}^{2}+q^{2})J_{1}(\eta_{l}R)J_{1}(\eta_{t}R)/R-
(\eta_{t}^{2}-q^{2})^{2}J_{0}(\eta_{l}R)J_{1}(\eta_{t}R)-4q^{2}\eta_{l}\eta_{t}J_{1}(\eta_{l}R)J_{0}(\eta_{t}R)=0$,
where $J_{0,1}(r)$ are Bessels of 1st kind, $R=D/2$ is the radius,
and $\eta_{l}=\sqrt{\omega^{2}/v_{l}^{2}-q^{2}}$, $\eta_{t}=\sqrt{\omega^{2}/v_{t}^{2}-q^{2}}$
may be interpreted as transverse wave vector components.
For fixed $\omega$, $D$, the dispersion relation possesses multiple solutions
$q_i = \omega/v_{q_i}(D),\, i=1,2,\ldots$. Each mode branch defines its own
phase velocity $v_{q_i}$, $v_{q_0} < v_{q_1} < \dots$ with the fundamental mode branch having
the smallest velocity.

The modes in the mpDBR cavity can be  constructed
as standing waves
using waveguide solutions for compressional modes:
\begin{eqnarray}
u_{r}(\bm{r}) & = & -A \eta_{l}\!\left[J_{1}(\eta_{l}r)+\frac{2q^{2}}{\eta_{t}^{2}-q^{2}}\frac{J_{1}(\eta_{l}R)}{J_{1}(\eta_{t}R)}J_{1}(\eta_{t}r)\right]\!\sin{qz}
\nonumber\\    
u_{z}(\bm{r}) & = & A q\left[J_{0}(\eta_{l}r)-\frac{2\eta_{l}\eta_{t}}{\eta_{t}^{2}-q^{2}}\frac{J_{1}(\eta_{l}R)}{J_{1}(\eta_{t}R)}J_{0}(\eta_{t}r)\right]\cos{qz}
\nonumber .    
\end{eqnarray}
The relevant strain components have $\sim \sin{qz}$ dependence,
i.e., the strain has a node at the Si-cavity $z$-boundaries
to ensure non-zero coupling with the donor $A_1 \leftrightarrow T_2$ transition
(see Eq.~2 of the main text).

The  constant $A$ can be calculated from a proper normalization condition:
\begin{equation}
\int_{V_{\rm cav}} d^3 \bm{r}\ \bm{u}^*_{\bm{q}}(\bm{r}) \bm{u}_{\bm{q}}(\bm{r}) = \gamma\, \frac{\hbar}{2 \rho \omega_{\bm{q}}}
\label{normalization} ,
\tag{S7}
\end{equation}
where $\bm{u}_{\bm{q}}(\bm{r}) = \bm{e}_r u_r(r,z) + \bm{e}_z u_z(r,z)$ is the displacement ``vacuum field''
(see Eq. \ref{displacement}) in the Si-cavity.
Here $\gamma$ is the fraction of the phonon vacuum energy, $\hbar\omega_{\bm{q}}/2$, stored in the cavity
of length $d_{c}$,
while an energy fraction $1-\gamma$ is stored in the DBR mirrors,
related to the penetration dept $\delta_p$ shown on Fig.~\ref{fig:1Dcavity}.
Approximate transfer matrix calculations based on the propagation of
four fields (the displacements $u_{z}$, $u_{r}$, and the stresses
$T_{zz}$, $T_{zr}$) along the mpDBR waveguide give  $\gamma \approx 0.3$
(note that the coupling $g$ scales as $\sim \sqrt{\gamma}$ which renormalizes the $g_{max}$'s calculated in the text)
that is commensurate with the value estimated from the
1D DBR field distribution.

For  each mode of mpDBR considered in resonance with the donor, a suitable
cavity design with appropriate cavity design parameters, i.e. cavity
and SL layer thicknesses, strain matched substrate Ge content, etc.
is required.
In accordance with our 1D DBR design, layer materials $A$ and $B$ are kept the same.
For a diameter of $D=\lambda_{l} = 12.3\, \mbox{nm}$,
a Si-cavity thickness of $d_{c} = \lambda_{q_0} = 6.9\, \mbox{nm}$  results with the fundamental
mode being in resonance with the P donor
(the layer thicknesses are correspondingly
$d_{A}=1.7\, \mbox{nm}$ and $d_{B}=1.2\, \mbox{nm}$).
Using the TM method, the ideal
$Q$-factor of the fundamental mode is calculated to range
from $10^{4} - 10^{6} $ for $N=20-33$ unit layers of the  DBR,
and similar numbers appear for the higher excited modes.

\subsection*{Anharmonicity and scattering off impurities}

In isotropic crystals at low temperatures ($T\ll\hbar\omega/k_{B}$)
a TA-phonon decay is forbidden, 
while the LA-phonon decays through $\mbox{LA}\to\mbox{TA}+\mbox{TA}$
and $\mbox{LA}\to\mbox{TA}+\mbox{LA}$ channels. For Si,
$\Gamma_{\rm LA}^{\rm tot}\approx 4.5\,10^{4}\,\nu_{\rm THz}^{5}$
\cite{Maris:1993p1295};
this gives for 3 meV ($0.73$ THz) a rate $1.4\,10^{4}\,\mbox{s}^{-1}$
(Si anisotropy lifts out the TA selection rule, leading to same order of magnitude TA-decay
rate; see, however, \cite{Maris:1993p1295}).
For larger temperatures
($T\gtrsim {3\mbox{meV}}/{k_{B}}\sim30\,\mbox{K}$), $\Gamma_{\rm LA}$
scales as $\omega T^{4}$ ({Landau-Rumer} effect, see e.g., Ref. \cite{Clelandbook2003})
leading to an
order of magnitude enhancement.
Stronger phonon losses come from scattering from isotopic point defects. For
natural Si the isotope mass fluctuations lead to a rate $\Gamma_{\rm imp} \approx 2.43\,10^{6} \nu_{\rm THz}^{4}$
\cite{Maris:1993p1295}, which gives $\sim 7\,10^{5}\,\mbox{s}^{-1}$ for $0.73\, \mbox{THz}$.
The estimated rate gives a phonon
mean free path of the order of $v_{l}/\Gamma_{\rm imp}\sim 1.2\,\mbox{cm}$.
An enrichment of $^{28}\mbox{Si}$ to 99\% (natural abundance of $^{28}\mbox{Si}$ is $92.23$\%)
decreases the scattering rate by one order of magnitude;
so, the dominating phonon loss may become scattering off the surfaces
and interfaces of the phonon microcavity \cite{Trigo:2006p923}.

In Ref. \cite{PascualWinter:2007p924} via direct measurement of the
phonon decay rate in a GaInAs SL cavity ($\nu_{1}=0.46\,\mbox{THz}$):
$\Gamma_{{\rm ph}}\simeq\Gamma_{{\rm conf}}+\Gamma_{{\rm anh}}$,
an anharmonicity rate was estimated as $\Gamma_{{\rm anh}}\simeq 3.3\,10^{9}s^{-1}$,
taking into account the phonon confinement due to the SL DBRs.
The observed large rates are nevertheless in qualitative agreement
with the above theoretical estimations, since the experiment was performed
at a high (room) temperature and dealt with large number of initially
injected phonons \cite{Clelandbook2003}. Thus, we expect that in
quantum-limited experiments at low temperatures ($\sim 1\,\mbox{K}$)
and low phonon occupation numbers  scattering rates
of the order of $10^{4}-10^{5}\,\mbox{s}^{-1}$ can be achieved.


\begin{thebibliography}{99}
\bibitem{zero-strain}
The intervalley transition $1s(A_{1})\to 1s(T_{2})$    
gives an odd in $z$ contribution, Eq. (\ref{matrix-elements-coupling}),
that overlaps with the strain,    
since at maximum $\bm{u}(\bm{r})$ strain changes sign.
\bibitem{1D-DBR} For 1D DBR cavity we evaluate:
$Q_{DBR} \simeq 2\pi(d_{c} + L_{DBR})\sqrt{R_c}/\lambda\ln{R_c}$,
where
$L_{DBR} = Z_A Z_B \lambda_q/2 Z_c |Z_B-Z_A|$ is the acoustic DBR mirror length
and
$R_{c}=\left[(Z_s/Z_c - Z_{r}^{2N})/(Z_s/Z_c + Z_{r}^{2N})\right]^{2}$
is the   
peak power reflectivity   
for $N$ layers with   
impedance ratio
$Z_{r} = \rho_A v_A/\rho_B v_B$
($Z_s/Z_c$ are the impedances of the substrate/cavity respectively).
\end{thebibliography}
\end{document}